\documentclass{article}

\PassOptionsToPackage{numbers, sort, compress}{natbib}
\usepackage[preprint,nonatbib]{nips_2018}

\usepackage[utf8]{inputenc} % allow utf-8 input
\usepackage[T1]{fontenc}    % use 8-bit T1 fonts
\usepackage{hyperref}       % hyperlinks
\usepackage{url}            % simple URL typesetting
\usepackage{booktabs}       % professional-quality tables
\usepackage{amsfonts}       % blackboard math symbols
\usepackage{nicefrac}       % compact symbols for 1/2, etc.
\usepackage{microtype}      % microtypography
\usepackage{graphicx}
\usepackage{subcaption}
\usepackage{amsthm}
\usepackage{amsmath}
\usepackage{algorithm}
\usepackage{algorithmic}
\usepackage{multirow}

\title{Learning to Learn Unlearned Feature for Brain Tumor Segmentation}

\author{
  Seungyub Han\thanks{Work performed while at Hodoo AI}, \hspace{0.1cm} Yeongmo Kim,  \hspace{0.1cm} Seokhyeon Ha,  \hspace{0.1cm} Jungwoo Lee,  \hspace{0.1cm} Seunghong Choi \textsuperscript{$\dagger$} \\
  Electrical and Computer Engineering,   Department of Radiology College of Medicine \textsuperscript{$\dagger$} \\
  Seoul National University\\
  \texttt{$\{$syhan, ymkim, hash1108$\}$@cml.snu.ac.kr, junglee@snu.ac.kr, verocay@snuh.org} \\
}

\begin{document}

\maketitle

\begin{abstract}
  We propose a fine-tuning algorithm for brain tumor segmentation that needs only a few data samples and helps networks not to forget the original tasks. Our approach is based on active learning and meta-learning.
  One of the difficulties in medical image segmentation is the lack of datasets with proper annotations, because it requires doctors to tag reliable annotation and there are many variants of a disease, such as glioma and brain metastasis, which are the different types of brain tumor and have different structural features in MR images. Therefore, it is impossible to produce the large-scale medical image datasets for all types of diseases. In this paper, we show a transfer learning method from high grade glioma to brain metastasis, and demonstrate that the proposed algorithm achieves balanced parameters for both glioma and brain metastasis domains within a few steps.
  
\end{abstract}

\section{Introduction}

The performance of semantic segmentation using deep neural networks has been improved recently. These segmentation networks are applied to the medical image analysis to help doctors to save time in diagnosis. The state-of-the-art networks still require large amounts of training data points for pre-training and fine-tuning. Gathering medical image datasets, however, is an expensive and time-consuming so that there are fewer datasets than the datasets for common objects \cite{pascal-voc-2011, DBLP:journals/corr/LinMBHPRDZ14, zhou2017scene}. In particular, in brain tumor segmentation, there is a proper dataset called BRaTS for the High Grade Glioma (HGG) and the Low Grade Glioma (LGG) \cite{menze2015multimodal}, and no well-annotated dataset for brain metastasis.

 HGG and brain metastasis have the different structural feature but the similar contrast feature. In contrast enhanced MR image, the region of brain tumor is highlighted because of the contrast media. In both cases, tumors have the similar contrast. These brain tumors have different pathological properties, so they have different structural characteristics. Therefore, a pre-trained network using the HGG dataset can not generate perfect segmentation for brain metastasis. In this paper, We learn the unlearned feature of brain metastasis without forgetting the pre-trained feature in order to optimize the balanced parameters between HGG and metastasis.

We first pre-trained the fully convolutional network (FCN) \cite{Long_2015_CVPR} with HGG in the dataset \cite{menze2015multimodal}. The gradient descent based fine tuning, which we call naive tune, uses many selected data points which have balanced instances per each class to produce the optimal fine-tuning results \cite{DBLP:journals/corr/LinMBHPRDZ14, zhou2017scene}. We propose two novel fine-tuning methods, passive meta-tune and active meta-tune, to optimize the pre-trained network. These two methods decide which training data points are first learned and update the network with \cite{finn2017model}. The orders of training dataset is determined with two active learning based rules, the passive learning and the active learning \cite{pmlr-v80-mussmann18a}. In this work, we produce the annotated brain metastasis data samples with 30 patients. Similar to the BraTS dataset, for a patient, there are 4 MR sequences, FLAIR, T1, T1 contrast-enhanced, T2, and 25 slices per each sequence.

\section{Proposed methodology}
\label{prop_model}
\begin{figure}[t]
    \centering
    \begin{subfigure}[b]{0.4\textwidth}
    \centering
        \includegraphics[width= \textwidth]{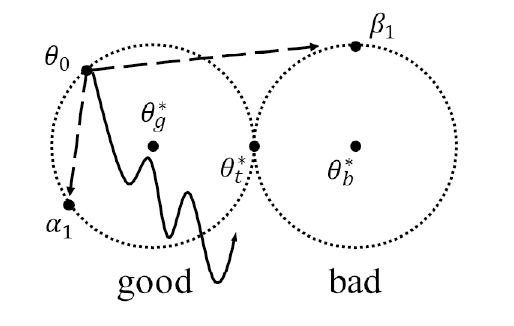}
        \caption{passive meta-tune method}
        \label{fig:passive}
    \end{subfigure}
    ~ %add desired spacing between images, e. g. ~, \quad, \qquad, \hfill etc. 
    \begin{subfigure}[b]{0.4\textwidth}
    \centering
        \includegraphics[width= \textwidth]{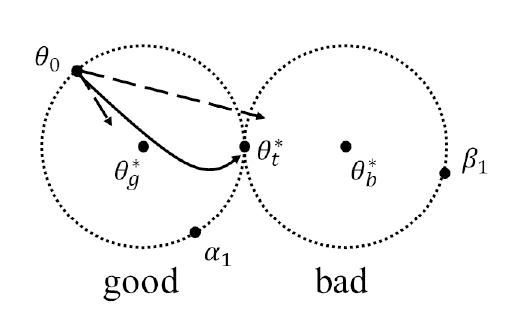}
        \caption{active meta-tune method}
        \label{fig:active}
    \end{subfigure}
    \caption{Two active learning based meta-tune methods : divide the results of segmentation task into the good result tasks (learned) and the bad result task (unlearned). At the first meta-tune step, the pre-trained parameter $\theta_0$ usually lies on the sub-optimal region of the pre-training optimum $\theta_{g}^*$. To reach the balanced target parameter $\theta_{t}^*$, sample the order of meta-tune inputs by each method. (a) is passive meta-tune that trains with randomly ordered inputs. The optimal parameters $\alpha_1, \beta_1$ for randomly sampled tasks lie in each circle. The update direction depends on the combination of $\alpha_i, \beta_i$. (b) is DSC based order inputs. The first input is the farthest data points across the optimums $\theta_{g}^* , \theta_{b}^*$. The induced update direction flows smoothly into $\theta_{t}^*$.}
    \label{fig:modelss}
\end{figure}
We propose two active learning based meta-tune methods, each of which is based on random sampling and a variant of uncertainty sampling \cite{pmlr-v80-mussmann18a}. We define the meta-tune as a fine-tune method using MAML algorithm \cite{finn2017model}. The meta-tune generalize the unlearned training examples more quickly than the graident-based fined tune method (naive tune). As shown in Figure (\ref{fig:modelss}), we train the model with the learned features continuously as well as the unlearned features not to forget the learned features.
\begin{algorithm}[h]
\caption{Active learning based meta-tune algorithms} \label{alg}
\begin{algorithmic}[1]
%\vspace{0.5mm}
\STATE Set learning rate hyper parameters : $\alpha, \beta$, pre-trained parameters : $\theta$ 
\STATE Divide inputs into inner-loop data and outer-loop data, Order inputs by passive or active method
\FOR {the meta batch size of one patient}
    \STATE Decompose tasks into good task $\mathcal{T}_{g}$, bad tasks $\mathcal{T}_{b}$
    \FOR {$\mathcal{T}_{i} \in$ $\mathcal{T}_{g} \cup \mathcal{T}_{b}$}
        \STATE Sample one data point $D = (\mathbf{x}, \mathbf{y})$ by each sample method
        \STATE Compute $\theta_{i}'=\theta-\alpha \nabla_{\theta}\mathcal{L}_{\mathcal{T}_i}(f_{\theta})$
        \STATE Sample one data point $D_{i}' = (\mathbf{x}', \mathbf{y}')$ by each sample method
    \ENDFOR
    \STATE Meta update $\theta = \theta - \beta \nabla_{\theta} \sum_{\mathcal{T}_i} \mathcal{L}_{\mathcal{T}_i}(f_{\theta_{i}'})$ using $D_i$
\ENDFOR
\end{algorithmic}
\end{algorithm}
\vspace{-0.5cm}

\section{Experimental Results}
\label{result}
 We clinically-acquire multimodal MRI scans and produce all the ground truth annotations by neuroradiologists. We use a VGG-16 based FCN network as the pre-train network. Then, we test our algorithms. (yellow: edema, green: necrosis, brown: ehnacing tumor: brown, red : high probability).
\begin{figure}[h]

\label{fig:model}
\centering
    \begin{subfigure}[b]{0.5\textwidth}
    \centering
        \includegraphics[scale=0.5]{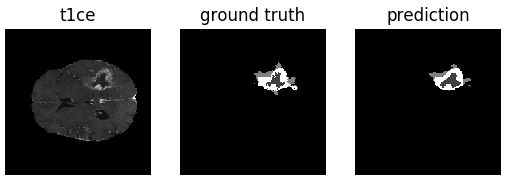}
        \caption{Pre-trained result on BraTS dataset}
        \label{fig:pretrain}
    \end{subfigure}
    ~ %add desired spacing between images, e. g. ~, \quad, \qquad, \hfill etc. 
    \begin{subtable}[b]{0.4\textwidth}
    \centering
        \label{res_table}
        \centering
  \begin{tabular}{lll}
    \toprule
    Method     & Dice Score ($\pm$std) \\
    \midrule
    baseline    & 0.66 (on BRaTS)    \\
    naive       & 0.33 $\pm$ 0.3413  \\
    passive     & 0.41 $\pm$ 0.2752  \\
    active      & \textbf{0.45 $\pm$ 0.2317}  \\
    \bottomrule
  \end{tabular}
    \caption{DSC result for enhaning tumor}
    \end{subtable}
\end{figure}

\begin{figure}[h]
    \centering
    \begin{subfigure}[h]{    \linewidth}
        \centering
        \includegraphics[width=     \linewidth]{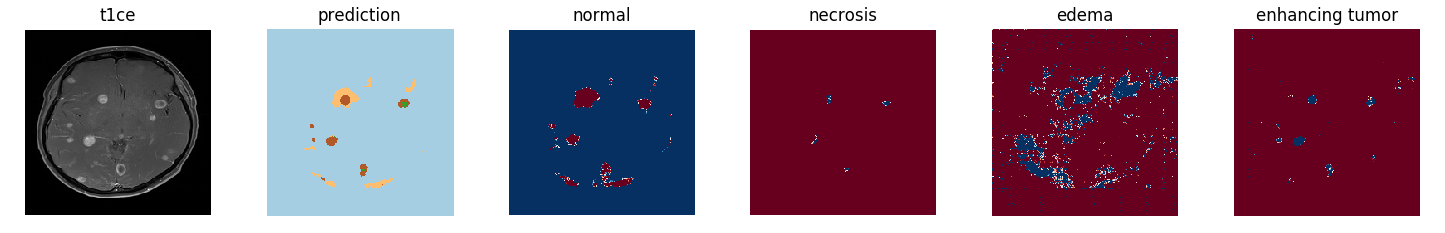}
        \caption{naive tune: all small tumors is detected, skull is misclassified as edema}
        \label{fig:gull}
    \end{subfigure}
    \begin{subfigure}[h]{     \linewidth}
        \centering
        \includegraphics[width=     \linewidth]{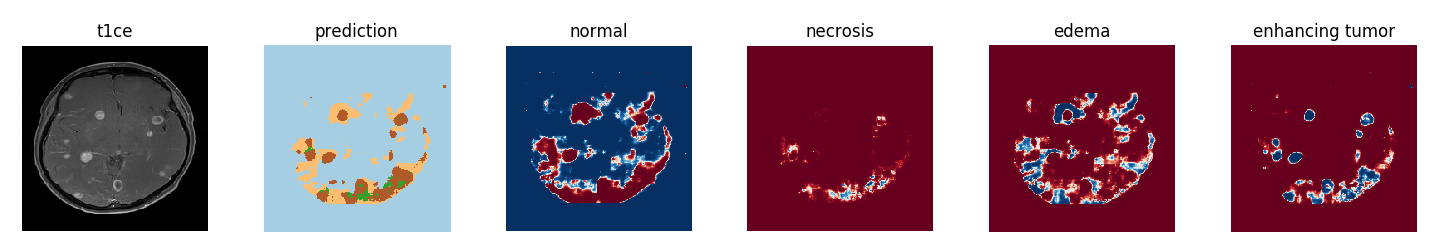}
        \caption{passive meta-tune: the region of edema is overestimated. this region invades the part of skull}
        \label{fig:gull}
    \end{subfigure}
    \begin{subfigure}[h]{     \linewidth}
        \centering
        \includegraphics[width=     \linewidth]{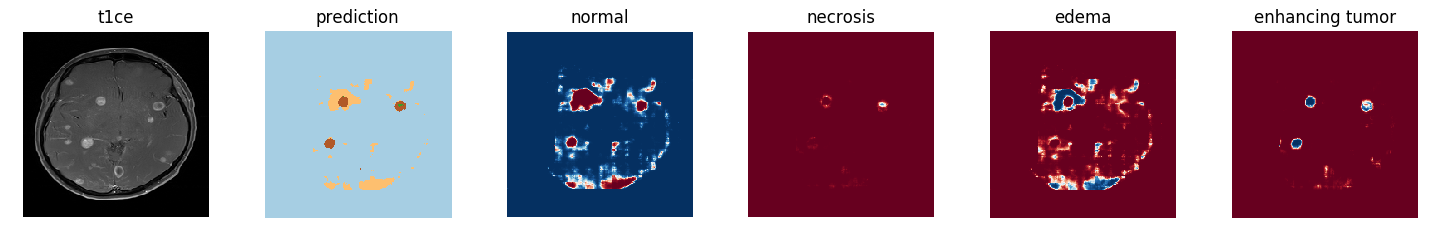}
        \caption{active meta-tune show the minimum false positive area}
        \label{fig:tiger}
    \end{subfigure}

  %  \caption{A segmentation result on the training dataset. the result of passive meta-tune oscillates.}\label{fig:animals}

    \begin{subfigure}[b]{     \linewidth}
        \centering
        \includegraphics[width=     \linewidth]{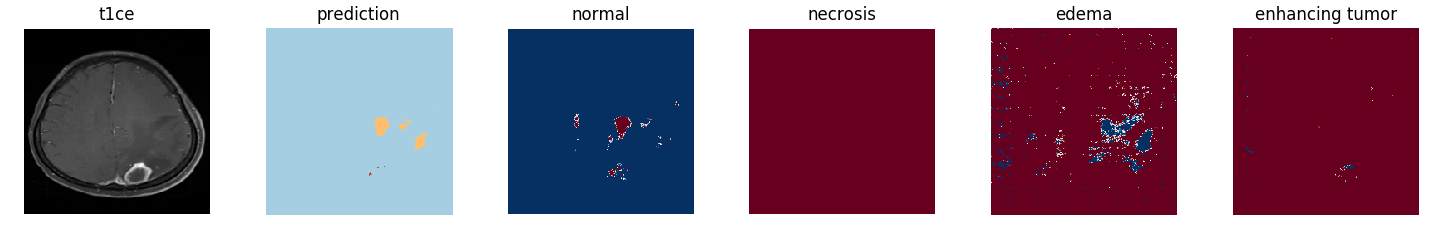}
        \caption{naive tune: unlike the previous case, a few step updates have no generalization effect}
        \label{fig:gull}
    \end{subfigure}
    \begin{subfigure}[b]{     \linewidth}
        \centering
        \includegraphics[width=     \linewidth]{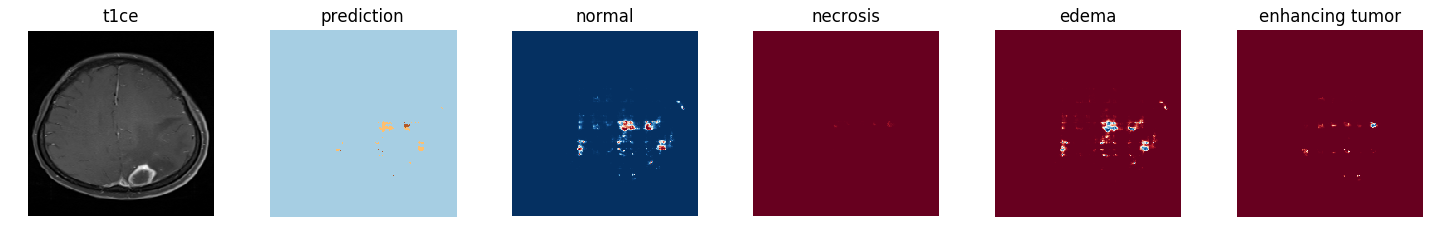}
        \caption{passive meta-tune: the worst case, which cannot capture a single tumor}
        \label{fig:tiger}
    \end{subfigure}
    \begin{subfigure}[b]{     \linewidth}
        \centering
        \includegraphics[width=     \linewidth]{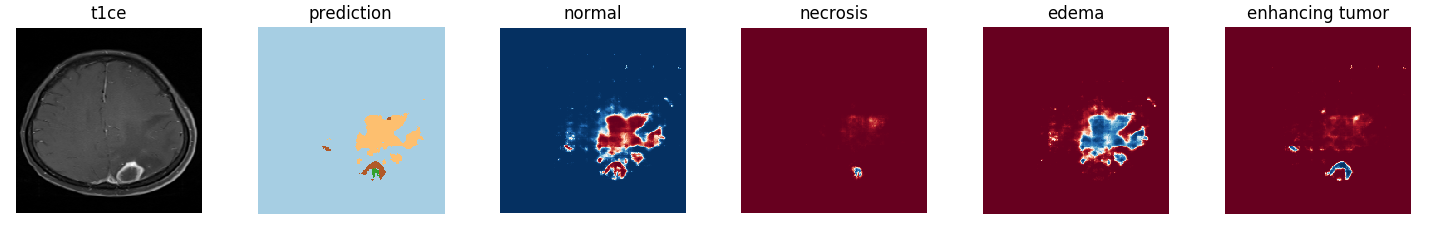}
        \caption{active meta-tune: the balanced optimal result}
        \label{fig:gull}
    \end{subfigure}
    \caption{Segmentation results on the training dataset (a)-(c). The result of passive meta-tune oscillates (b). Segmentation result on the validation dataset (d)-(f).}
\end{figure}
\section{Conclusion}
\label{conclusion}
We proposed an active meta-tune method which learns unlearned feature without forgetting the original task. We show that our method have a generalization effect within the target domain segmentation (brain metastasis).We expect our method can be extended to other medical lesion applications.
\clearpage

\subsubsection*{Acknowledgments}
This work is in part supported by SNU Eng-Med Collaboration Grant, Basic Science Research Program (NRF-2017R1A2B2007102) through NRF funded by MSIP, Technology Innovation Program (10051928) funded by MOTIE, Bio-Mimetic Robot Research Center funded by DAPA (UD130070ID), INMAC, and BK21-plus.
\bibliography{main}
\bibliographystyle{plain}

\end{document}